\documentstyle[aps,eqsecnum]{revtex}

\newcommand{\be}{\begin{equation}}
\newcommand{\ee}{\end{equation}}
\newcommand{\bea}{\begin{eqnarray}}
\newcommand{\eea}{\end{eqnarray}}
\newcommand{\nn}{\nonumber}

\draft

\begin{document}

\author{Arsen Khvedelidze~$^{a,b}$ ~~and~~ Dimitar Mladenov $^b$}

\title
{Euler-Calogero-Moser system from $SU(2)$ Yang-Mills theory}

\address{$^a$~Department of Theoretical Physics,
A.Razmadze Mathematical Institute, GE-380093 Tbilisi, Georgia}

\address{$^b$~Bogoliubov Laboratory of Theoretical Physics,
Joint Institute for Nuclear Research, Dubna, Russia}

\date{\empty}

\maketitle

\begin{abstract}
The relation between $SU(2)$ Yang-Mills mechanics, originated from the 4-dimensional
$SU(2)$ Yang-Mills theory under the supposition of spatial homogeneity of the gauge
fields, and the Euler-Calogero-Moser model is discussed in the framework of Hamiltonian
reduction.
Two kinds of reductions of the degrees of freedom are considered:
due to the gauge invariance and
due to the discrete symmetry.
In the former case, it is shown that after elimination of the gauge degrees of freedom
from the $SU(2)$ Yang-Mills mechanics the resulting unconstrained system
represents the ${\rm ID_3}$ Euler-Calogero-Moser model
with an external fourth-order potential.
Whereas in the latter, the ${\rm IA_6}$ Euler-Calogero-Moser model
embedded in an external potential is derived
whose projection onto the invariant submanifold through the discrete
symmetry coincides again with the $SU(2)$ Yang-Mills mechanics.
Based on this connection, the equations of motion
of the $SU(2)$ Yang-Mills mechanics in the limit of the zero coupling constant
are presented in the Lax form.\\[0.5cm]

\end{abstract}



\section{Introduction}


The present note is devoted to the discussion of the correspondence
between the dynamics of $3$-particles with internal degrees interacting by pairwise
$1/r^3 $ forces on a line (Euler-Calogero-Moser system \cite{GibbonsHermsen,Wojciechowski})
and $SU(2)$ Yang-Mills theory with spatially constant gauge fields
($SU(2)$ Yang-Mills mechanics \cite{MatSav}
(see also  \cite{YMM,DYMM} and references therein)).

The Euler-Calogero-Moser model is the extension of the famous
Calogero-Sutherland-Moser models \cite{Calogero,Sutherland,Moser}
(for its generalizations see \cite{Perelomov}
and reviews \cite{OlshanetskyPerelomov,PerelomovBook})
with additional dynamical internal degrees of freedom included.
It is interesting that these types of models arises in various areas of theoretical
physics like
the 2-dimensional Yang-Mills theory \cite{D2YangMills},
black hole physics \cite{GibbonsTownsend},
spin chain systems \cite{SpinChain},
generalized statistics \cite{GenStatistics},
higher spin theories \cite{MVasiliev},
level dynamics for quantum systems \cite{LevelDynamics},
quantum Hall effect \cite{QHall}
and many others.
An attractive feature of these generalizations is that
it maintains the integrability property of the original Calogero-Sutherland-Moser system.
For the general elliptic version of the Euler-Calogero-Moser system,
the action-angle type variables have been constructed and the equations of
motion have been solved in terms of Riemannian theta-functions \cite{KriBaBiTa},
the canonical symplectic form of this model is represented
in terms of algebro-geometric data \cite{BabelonTalon}
using the general construction of Krichever and Phong \cite{KricheverPhong}.

During the past years a remarkable relation between the Calogero-Moser systems and the exact
solutions of $4$-dimensional supersymmetric gauge theories
has been found \cite{SeibergWitten}.
It has been recognized  that the so-called Seiberg-Witten spectral curves
are identical to the spectral curves of
the elliptic $SU(N)$ Calogero-Moser system \cite{SpectralCurves}.
Furthermore the generalization of these relations to the
$N=2$ supersymmetric gauge theories with general Lie algebras
and an adjoint representation of matter hypermultiplet have been
derived in \cite{D'HokerPhong}
(for review of the recent results see, e.g., \cite{Review}).

Despite the existence of such a correspondence established on very
general grounds, relations between gauge theories and integrable models are far from
being understood.
In the present note, we would like to point out a simple direct correspondence
between the $SU(2)$ Yang-Mills theory and the Euler-Calogero-Moser model.
This correspondence follows from the sequence of reductions of degrees
of freedom thanks to different kinds of symmetries.
At first, supposing the spatial homogeneity of gauge fields, the field
theory is reduced to the 9-dimensional degenerate Lagrangian model.
Then the pure gauge variables are eliminated by applying the method of
Hamiltonian reduction.
Finally, rewriting the derived unconstrained matrix model
in terms of special coordinates adapted to the action of rigid symmetry,
one can arrive at the conventional form of the Euler-Calogero-Moser Hamiltonian.
More precisely, we shall demonstrate that the unconstrained $SU(2)$ Yang-Mills
mechanics represents the Euler-Calogero-Moser system of type ${\rm ID_3}$,
i.e., the inverse-square interacting 3-particle system with internal degrees of freedom
related to the root system of simple Lie algebra
$D_3$ \cite{OlshanetskyPerelomov,PerelomovBook},
and is embedded in a fourth order external potential written in the superpotential form.

Besides this reduction due to the continuous symmetry of the system,
we discuss another possibility of relating the Yang-Mills
mechanics to higher order matrix models using the discrete symmetries.
We shall explore the method of constructing generalizations of the
Calogero-Sutherland-Moser models elaborated recently by
A. Polychronakos \cite{PolyDiscrete}.
This method consists in the usage of the appropriate reduction of
the original Calogero model by a subset of its discrete symmetries
to an invariant submanifold of the phase space.
Representing the Euler-Calogero-Moser system with a special external potential
as a $6\times 6$ symmetric matrix model, we shall show that such a matrix model,
after projection onto the invariant submanifold of the phase space using a
certain subset of discrete symmetries, is equivalent to the unconstrained
$SU(2)$ Yang-Mills mechanics.
Finally, we give a Lax pair representation for the equations of motion of
the $SU(2)$ Yang-Mills mechanics in the limit of the zero coupling constant.


\section{Hamiltonian reduction of the Yang-Mills mechanics}



\subsection{The equivalent unconstrained matrix model}


The dynamics of the $SU(2)$ Yang-Mills 1-form ${\bf A}$ in 4-dimensional Minkowski
space-time $M_4$ is governed by the conventional local functional
\be \label{eq:YMA}
S_{YM}  =  \frac{1}{2} \int_{M_4} \mbox{tr}\, F \wedge \ast F \,,
\ee
defined in terms of the curvature 2-form
$
F = d {\bf A}  + g\,{\bf A} \wedge {\bf A} \,,
$
with the coupling constant $g$.
After the supposition of the spatial homogeneity of the connection ${\bf A}$
\be \label{eq:sph}
{\cal L}_{\partial_i} {\bf A} = 0 \,,
\ee
the action (\ref{eq:YMA}) reduces to the action for a finite dimensional model, the so-called
Yang-Mills mechanics (YMM) described by the degenerate matrix Lagrangian
\be
L_{YMM} = \frac{1}{2} \mbox{tr} \left( (D_t A) ( D_t A)^T \right) - V(A)\,,
\ee
The entries of the $3\times 3 $ matrix $A$ are nine spatial components
$A_{ai} := A^{a}_{i}$ of the connection
$
{\bf A} := Y_a e_a dt + A_{ai} e_a dx^i \,,
$
where
$e_a = \sigma_a/2 i$ with the Pauli matrices $\sigma_a$ and
$D_t$ denotes the covariant derivative
$
(D_t A)_{ai} = \dot{A}_{ai} + g \varepsilon_{abc} Y_b A_{ci}\,.
$
Due to the spatial homogeneity condition (\ref{eq:sph}),
all dynamical variables  $Y_a$ and $A_{ai}$ are functions
of time  only.
The part of the Lagrangian  corresponding to the self-interaction
of the gauge fields is gathered in the potential $V(A)$
\be
V(A)  = \frac{g^2}{4}
\left(
\mbox{tr}{}^2 ( A A^T ) -  \mbox{tr} ( A A^T )^2
\right) \,.
\ee
To express the Yang-Mills mechanics in a Hamiltonian form, let us define
the phase space endowed with the canonical symplectic structure and
spanned by the canonical variables $ (Y_a,  P_{Y_a}) $
and $ (A_{ai}, E_{ai}) $
where
\bea \label{eq:canm}
P_{Y_a}  =  \frac{\partial L}{\partial \dot{Y}_a} = 0\,, \quad
E_{ai}   =  \frac{\partial L}{\partial \dot{A}_{ai}} =
\dot{A}_{ai} + g\varepsilon_{abc} Y_b A_{ci}\,.
\eea
According to these definitions of the canonical momenta (\ref{eq:canm}),
the phase space is restricted by the three primary constraints
\be  \label{eq:boscon}
P^a_Y = 0\,
\ee
and the evolution of the system is governed by the total Hamiltonian
$
H_T = H_C + u^a_Y (t) \, P^a_Y \,,
$
where the canonical Hamiltonian is given by
\be
H_C = \frac{1}{2} \mbox{tr}(E E^T)  +
\frac{g^2}{4}
\left(
\mbox{tr}{}^2 ( A A^T ) -  \mbox{tr} ( A A^T )^2
\right) + g Y_a\,\mbox{tr}(J_a A E^T)\,,
\ee
and the matrix $(J_a)_{bc}$ is defined by $(J_a)_{bc} =\,-\varepsilon_{abc}$.
The conservation of constraints (\ref{eq:boscon})
in time entails the further condition on the canonical variables
\be  \label{eq:secconstr}
\Phi_a  = g\,\mbox{tr}(J_a A E^T) = 0\,,
\ee
that reproduces the homogeneous part of the conventional non-Abelian Gauss law constraints.
They are the first class constraints obeying the  Poisson brackets algebra
\be  \label{eq:algb1}
\{ \Phi_a , \Phi_b \} =  \varepsilon_{abc}\Phi_c \,.
\ee
In order to project onto the reduced phase space, we use
the well-known polar decomposition for an arbitrary $3\times3$ matrix
\be \label{eq:pcantr}
A_{ai}(\phi, Q) = O_{ak}( \phi ) Q_{ki}\,,
\ee
where $Q_{ij} $ is a positive definite $3\times3$ symmetric matrix and
$O(\phi_1,\phi_2, \phi_3) = e^{\phi_1 J_3}e^{\phi_2 J_1}e^{\phi_3 J_3}$
is an orthogonal matrix $O \in SO(3)$.
Assuming the nondegenerate character of the matrix $A_{ai}$,
we can treat the polar decomposition as
uniquely invertible transformation from the configuration variables
$A_{ai}$ to a new set of six Lagrangian coordinates $Q_{ij}$ and three coordinates $\phi_i$.
As it follows from further consideration, the variables parameterizing
the elements of the $SO(3)$ group (Euler angles $(\phi_1, \phi_2, \phi_3)$)
are the pure gauge degrees of freedom.

The field strength $E_{ai}$ in terms of the new canonical variables is
\be  \label{eq:potn}
E_{ai} = O_{ak}(\phi)
\biggl[
\,  P_{ki} + \varepsilon _{kil} (\gamma^{-1})_{lj}
\left[
\xi^L_j  - S_j                             \,
\right]\,
\biggr]\,,
\ee
where $\xi^L_a$ are three left-invariant vector fields on $SO(3)$
\bea
&& \xi^L_1 =
\frac{ \sin\phi_3 }{ \sin\phi_2 }\,  P_1 +
\cos\phi_3 \,  P_2 -
\cot\phi_2 \sin\phi_3 \  P_3 \,,\\
&& \xi^L_2 =
\frac{ \cos\phi_3 }{ \sin\phi_2 }\,  P_1 -
\sin\phi_3 \,  P_2 -
\cot\phi_2 \cos\phi_3 \ P_3 \,,\\
&& \xi^L_3 =  P_3\,.
\eea
Here $S_j = \varepsilon_{jmn}( P Q )_{mn}$ is the spin vector of the gauge field
and
\be
\gamma_{ik} = Q_{ik} -  \delta_{ik} \ \mbox{tr}  Q\,.
\ee

Reformulation of the theory in terms of these variables allows one to easily achieve the
Abelianization of the secondary Gauss law constraints.
Using the representations (\ref{eq:pcantr}) and (\ref{eq:potn}),
one can convince oneself that the variables $ Q_{ij} $ and $P_{ij}$ make no contribution to the
secondary constraints (\ref{eq:secconstr})
\be \label{eq:gauss}
\Phi_a = O_{ab}(\phi) \, \xi^L_b = 0\,.
\ee
Hence, assuming nondegenerate character of the matrix
\be
M\, =
\left (
\begin{array}{ccc}
\frac{\sin\phi_1}{\sin\phi_2}\,, & \cos\phi_1\,, &-\sin\phi_1 \cot\phi_2\\
-\frac{\cos\phi_1}{\sin\phi_2}\,,& \sin\phi_1\,, &\cos\phi_1 \cot\phi_2\\
0\,,                             &     0\,,      &           1
\end{array}
\right )\,,
\ee
we find the set of Abelian constraints equivalent to the Gauss law (\ref{eq:secconstr})
\be
{\tilde\Phi}_a = P_a = 0 \,.
\ee
After having rewritten the model in this form, we are able
to reduce the theory to physical phase space by a straightforward projection
onto the constraint shell .
The resulting unconstrained Hamiltonian,
defined as a projection of the total Hamiltonian onto the constraint shell
\be
H_{YMM}:= H_C( Q_{ab}\,, P_{ab}) \, \Bigl\vert_{P_a = 0\,,\ P^a_Y = 0}\,,
\ee
can be written in terms of $Q_{ab}$ and $P_{ab}$ as
\be \label{eq:uncYMP}
H_{YMM} \, = \,
\frac{1}{2}  \mbox{tr} P^2 -
\frac{1}{ \det^2\gamma }\mbox{tr}\, (\gamma {\cal M}\gamma )^2 +
\frac{g^2}{4} \left(
\mbox{tr}{}^2 Q^2  -  \mbox{tr} Q^4
\right) \, ,
\ee
where ${\cal M}_{mn} = (QP-PQ)_{mn}$ denotes the gauge field spin tensor.


\subsection{Unconstrained model as particle motion on stratified manifold}


In the previous section, the unconstrained dynamics of the SU(2) Yang-Mills mechanics
was identified with the dynamics of the nondegenerate matrix model
(\ref{eq:uncYMP}).
The configuration space ${\cal Q }$ of the real symmetric $3\times 3$
matrices can be endowed with the flat Riemannian metric
\be
ds^2 = \mbox{Tr}\left(dQ^2\right)
\ee
whose group of isometry is formed by orthogonal transformations
\be
Q' = RQR^T
\ee
Since the unconstrained Hamiltonian system (\ref{eq:uncYMP})
is invariant under the action of this rigid group,
we are interested in the structure of the orbit space given as a quotient
${\cal Q }/SO(3)$.
The important information on the stratification of
the space ${\cal Q }/SO(3)$ of orbits can be obtained  from
the so-called isotropy group of points of configuration space
which is defined as a subgroup of $SO(3)$ leaving point $x$ invariant
$RxR^T=x$.
Orbits with the same isotropy group are collected into classes,
called by {\it strata}.
So, as for the case of symmetric matrix, the orbits are uniquely parameterized by the
set of ordered eigenvalues of the matrix
$Q$ $x_1\leq x_2 \leq x_3$.
One can classify the orbits according to the isotropy groups which are determined by the
degeneracies of the matrix eigenvalues:
\begin{enumerate}
\item
{\it Principal orbit-type strata},
when all eigenvalues are unequal $x_1< x_2 < x_3$ with
the smallest isotropy group $Z_2\otimes Z_2$\,.
\item
{\it Singular orbit type strata}
forming the boundaries of orbit space with
\begin{enumerate}
\item two coinciding eigenvalues
$x_1=x_2, x_2=x_3 $ or $ x_1=x_3 $,
the isotropy group is $SO(2)\otimes Z_2$\,.
\item all three eigenvalues are equal
$x_1=x_2 =x_3$,
here the isotropy group coinciding with the isometry group $SO(3)$.
\end{enumerate}
\end{enumerate}

In the subsequent sections, we shall demonstrate that the dynamics of the
Yang-Mills mechanics, which takes place on the {\it principal} orbits is
governed by the $\rm{ID}_3$ Euler-Calogero model Hamiltonian with the external potential
$V^{(3)}:= g^2/2\sum_{i<j}x^2_ix^2_j$,
while for singular orbits the corresponding system is either
the $A_2$ Calogero model with the external potential
$V^{(2)}:=g^2/2(x^4 + 2\,x^2y^2)$ for {\it singular} orbits of type (a)
or one dimensional system with quartic potential
$V^{(1)}:= 3/2g^2x^4$ for singular orbits of type (b).


\subsubsection{Hamiltonian on principal orbit strata}


To write down the Hamiltonian describing the motion on the principal orbit strata,
we introduce the coordinates along the slices $x_i$ and along the orbits $\chi$.
Namely, we decompose the nondegenerate symmetric matrix $ Q $ as
\be
Q = {\cal R}^T\, (\chi_1, \chi_2, \chi_3)\, {\cal D } \
{\cal{R}}(\chi_1, \chi_2, \chi_3)
\ee
with the $ SO(3) $ matrix  ${\cal R}$ parameterized by the three Euler angles
$\chi_i := (\chi_1, \chi_2, \chi_3 )$
and the diagonal matrix
${\cal D}  = \mbox{diag}\ ( x_1 , x_2 , x_3 )$
and consider it as point transformation from the
physical coordinates $ Q_{ab} $ and $ P_{ab} $ to
$x_i,p_i$ and $\chi_i, p_{\chi_i}$.
The Jacobian of this transformation is the relative volume of orbits
\be
 J:= \Biggl|
\det{
\biggl| \biggl|
 \frac{\partial Q}{\partial x_k}, \frac{\partial Q}{\partial \chi_k}
\biggr|\biggr|
}
\Biggr| =
\prod_{i<k} \mid x_i- x_k \mid
\ee
and is regular for this stratum $x_1< x_2 < x_3$.

By using the generating function
\be
F \left[ x_i, \chi_i; \ P \right]  =
\mbox{tr}\, \left(Q P \right) =
\mbox{tr}\, \left( {\cal R }^T( \chi ) {\cal D}(x) {\cal R }( \chi ) P \right)
\ee
the canonical conjugate momenta can be found in the form
\bea
p_i = \frac{\partial F}{\partial x_i} =
\,\mbox{tr}
\left(
P {\cal R}^T \overline{\alpha}_i {\cal R}
\right) \,,\quad \quad
p_{\chi_i} =  \frac{\partial F }{\partial \chi_i} =
\mbox{tr}
\left(
{\cal R}^T \frac{\partial {\cal R}}{\partial \chi_i} \,(P Q - Q P)
\right)\,,
\eea
where $\overline{\alpha}_i$ are the diagonal members of the orthogonal basis
for the symmetric $3 \times 3$ matrices
$\alpha_A = ( \overline{\alpha}_i ,\ \alpha_i ) \  i = 1, 2, 3 $
under the scalar product
\bea
\mbox{tr} (\bar\alpha_a\,, \bar\alpha_b)= \delta_{ab}\,,
\quad
\mbox{tr} (\alpha_a\,, \alpha_b)=2\delta_{ab}\,,
\quad
\mbox{tr} (\bar\alpha_a\,, \alpha_b)= 0\,.\nn
\eea
The original physical momenta $P_{ik}$ can then be expressed in terms of the new canonical
variables as
\bea \label{eq:newmom}
P  =
{\cal R }^T
\left(
\sum_{s=1}^3  \bar{\cal P}_s \, \overline{\alpha}_{s} +
\sum_{s=1}^3 {\cal P}_s \, {\alpha}_{s}\right) \, {\cal R}
\eea
with $\bar{\cal{P}}_s = p_s$,
\bea
{\cal P}_i   = - \frac{1}{2} {\frac{\xi^R_i}{ x_j - x_k }}, \,\,\,\,\,
(\mbox{cyclic}\,\,\,\, \mbox{permutation} \,\,\, i\not=j\not= k )
\eea
and the $SO(3)$ right-invariant Killing vectors
\bea
&& \xi^R_1 = p_{\chi_1}\,,\\
&& \xi^R_2 =
- \sin\chi_1 \cot\chi_2 \ p_{\chi_1} +
\cos\chi_1 \  p_{\chi_2}             +
\frac{\sin\chi_1}{\sin\chi_2}\ p_{\chi_3} \,,\\
&& \xi^R_3 =
\,\,\cos\chi_1 \cot\chi_2 \ p_{\chi_1}   +
\sin\chi_1 \  p_{\chi_2}             -
\frac{\cos\chi_1}{\sin\chi_2}\ p_{\chi_3} \,.
\eea
They satisfy the Poisson bracket algebra
\be
\{\xi^R_a, \xi^R_b\} = \varepsilon_{abc} \xi^R_c \,.
\ee
Thus, finally, we get the following physical Hamiltonian defined on
the unconstrained phase space
\be  \label{eq:PYM}
H_{YMM} =
\frac{1}{2} \sum_{a=1}^3  p^2_a +
\frac{1}{4}\sum_{a=1}^3 k_a^2 \xi^2_a + V^{(3)}(x) \,,
\ee
where
\be
k^2_a = \frac{1}{(x_b +  x_c)^2}  + \frac{1}{(x_b -  x_c)^2}\,,
\qquad \mbox{cyclic} \quad a\not=b\not=c
\ee
and
\be\label{eq:YMMpot}
V^{(3)} = \frac{g^2}{2} \  \sum_{a < b}  x_a^2 x_b^2\,.
\ee

Note  that the potential term in (\ref{eq:YMMpot}) has symmetry beyond the cyclic one.
This fact allows us to write $V^{(3)}(x_1, x_2, x_3)$ in the form
\be \label{suppot}
V^{(3)}(x_1, x_2, x_3) =
\frac{\partial W^{(3)} }{\partial x_a}\frac{ \partial W^{(3)} }{ \partial x_a }\,,
\qquad a = 1,\,2,\,3
\ee
with the superpotential $W^{(3)} = x_1 x_2 x_3$.

This completes our reduction of the spatially homogeneous
Yang-Mills system to the equivalent unconstrained system
describing the dynamics of the physical dynamical degrees of freedom.
We see that the reduced Hamiltonian $H_{YMM}$ on the principal orbit strata
is exactly the Hamiltonian of the Euler-Calogero-Moser system of type  $\rm{ID}_3$,
i.e., is of the inverse-square interacting 3-particle system with internal
degrees of freedom and related to the
root system of the simple Lie algebra $\rm {D}_3$
\cite{OlshanetskyPerelomov,PerelomovBook}
embedded in the fourth order external potential (\ref{suppot}).


\subsection{Singular stratum}


Introduction of the additional constraints
\be
x_1-x_2=0
\ee
or
\be
x_1-x_2=0\,, \quad \quad x_1-x_3=0
\ee
defines the invariant two- and one- dimensional strata.
One can repeat the above consideration for these singular strata and derive,
correspondingly, the following unconstrained Hamiltonians:


\subsubsection{Two-dimensional strata}


\be \label{eq:YMC}
H^{(2)}_{Sing} =
\frac{1}{2} (p^2_x + p^2_y) +
\frac{1}{4} \frac{l(l+1)}{(x-y)^2} + \frac{g^2}{2}(x^4 + 2\,x^2y^2) \,,
\ee
where the constant $l(l+1)$ denotes a value of the square of the particle
internal spin.


\subsubsection{One-dimensional strata}


\be \label{eq:YMC1}
H^{(1)}_{Sing} = \frac{1}{2} p^2_x + 3/2g^2x^4 \,.
\ee


\section
{Euler-Calogero-Moser system as a free motion on space of symmetric matrices}


In order to discuss the relation between the Yang-Mills mechanics and the
Euler-Calogero-Moser system, it is useful to represent the later in the form of
a nondegenerate matrix model.
Let us consider the Hamiltonian system with the phase space
spanned by the $N \times N$ symmetric matrices $X$ and $P$ with the
noncanonical symplectic form
\be
\{X_{ab}, P_{cd}\} = \frac{1}{2}
\left(
\delta_{ac} \delta_{bd} - \delta_{ad} \delta_{bc}
\right) \,.
\ee
The Hamiltonian of the system defined as
\be\label{eq:freematrix}
H = \frac{1}{2} \mbox{tr} P^2
\ee
describes a free motion in the matrix configuration space.
The following statement is fulfilled:\\
{\it The Hamiltonian (\ref{eq:freematrix}) rewritten in special coordinates
coincides with the Euler-Calogero-Moser Hamiltonian}
\be \label{eq:freeECM}
H = \frac{1}{2} \sum_{i=1}^N p_i^2 +
\frac{1}{2} \sum_{i\not=j}^N \frac{l_{ij}^2}{(x_i-x_j)^2} \,.
\ee
with nonvanishing Poisson brackets for the canonical variables
\footnote{
This system is the spin generalization of the Calogero-Moser model.
Particles are described by their coordinates $x_i$ and momenta $p_i$ together with internal
degrees of freedom of angular momentum type $l_{ij} = -\, l_{ji}$.
The analogous model has been introduced in \cite{GibbonsHermsen} where the
internal degrees of freedom satisfy the following Poisson brackets relations
\be
\{l_{ab}, l_{cd} \} = \delta_{bc} l_{ad} -\delta_{ad} l_{cb} \,.
\ee
}

\bea  \{x_i, p_j \} =
\delta_{ij} \, \quad \{l_{ab}, l_{cd} \} = \frac{1}{2} \left( \delta_{ac}
l_{bd} -\delta_{ad} l_{bc} + \delta_{bd} l_{ac} -\delta_{bc} l_{ad} \right)\,,
\eea

To find the adapted set of coordinates in which the Hamiltonian
(\ref{eq:freematrix})
coincides with the Euler-Calogero-Moser Hamiltonian (\ref{eq:freeECM}),
let us introduce new variables
\be
X = O^{-1}(\theta) Q(q) O(\theta)  \,,
\ee
where the orthogonal matrix $O(q)$ is parameterized by the
$\frac{N(N-1)}{2}$ elements, e.g., the Euler angles
$(\theta_1, \cdots , \theta_{\frac{N(N-1)}{2}})$ and
$Q = diag\|q_1, \cdots , q_N \|$ denotes a diagonal matrix.
This point transformation induces the canonical one which we can
obtain using the generating function
\be
F_4 = \left[
P, q_1, \cdots , q_N, \theta_1, \cdots , \theta_{\frac{N(N-1)}{2}}
\right] =\mbox{tr} [X(q,\theta) P]\,.
\ee
Using the representation
\be
P = O^{-1} \left[
\sum_{a=1}^N {\bar\alpha}_a {\bar P}_a  +
\sum_{i<j=1}^{\frac{N(N-1)}{2}} \alpha_{ij} P_{ij} \,
\right] O\,,
\ee
where the matrices $({\bar\alpha}_a, \alpha_{ij})$ form an orthogonal basis
in the space of the symmetric $N \times N$ matrices under the scalar product
\be
\mbox{tr} ({\bar\alpha}_a {\bar\alpha}_b ) = \delta_{ab}  \,,\qquad
\mbox{tr} (\alpha_{ij} \alpha_{kl} ) = 2 \delta_{ik} \delta_{jl} \,, \qquad
\mbox{tr} (\alpha_a \alpha_{ij}) =  0\,,
\ee
one can find that ${\bar P}_a = p_a$ and components $P_{ab}$ are represented
via the $O(N)$ right invariant vectors fields $l_{ab}$
\be
P_{ab} = \frac{1}{2} \frac{l_{ab}}{x_a - x_b} \,.
\ee
From this, it is clear that the Hamiltonian (\ref{eq:freematrix})
coincides with the Euler-Calogero-Moser Hamiltonian (\ref{eq:freeECM}).

The integration of the Hamilton equations of motion
\bea
\dot X  = P \,,\quad \quad
\dot P  = 0
\eea
derived with the help of Hamiltonian (\ref{eq:freematrix}),
gives the solution of the Euler-Calogero-Moser Hamiltonian system as follows:
for the $x$-coordinates we need to compute the eigenvalues of the matrix
$X = X(0) + P(0) t$, while the orthogonal matrix $O$,
which diagonalizes $X$, determines the time evolution of internal variables.


\section
{Yang-Mills mechanics through the discrete reduction of Euler-Calogero-Moser system}


In this section, we shall demonstrate how the $SU(2)$ Yang-Mills mechanics
arises from the higher dimensional matrix model after projection onto a
certain invariant submanifold determined by the discrete symmetries.
Let us consider the classical Hamiltonian system of $N$ particles on a line
with internal degrees of freedom embedded in external field with the
potential $V(x_1, x_2, \ldots, x_N)$ and described by the Hamiltonian
\be \label{eq:ECM}
H = \frac{1}{2} \sum_{i=1}^N p_i^2 +
\frac{1}{2} \sum_{i\not=j}^N \frac{l_{ij}^2}{(x_i-x_j)^2} +
V^{(N)}(x_1, x_2, \ldots, x_N) \,.
\ee
The particles are described by their coordinates $x_i$ and momenta $p_i$
together with the internal degrees of freedom of angular momentum type
$l_{ij} = -\, l_{ji}$.
The nonvanishing Poisson brackets are
\bea
\{x_i, p_j \} = \delta_{ij} \, \quad \quad
\{l_{ab}, l_{cd} \} = \delta_{ac} l_{bd} -\delta_{ad} l_{bc} +
\delta_{bd} l_{ac} - \delta_{bc} l_{ad} \,.
\eea
The external potential $V^{(N)}(x_1, x_2, \ldots, x_N) $
is constructed in terms of the superpotential $ W^{(N)} $
\be \label{pot6}
V^{(N)}(x_1, x_2, \ldots, x_N) =
-\frac{1}{4} \sum_{a=1}^N
\frac{\partial W^{(N)}}{\partial x_a} \frac{\partial W^{(N)}}{\partial x_a} \,,
\ee
with $W^{(N)}$ given as
\footnote{
Writing the superpotential in an invariant form
as
$$
W^{(N)} = i \sqrt{\mbox{det} X}\,,
$$
with the help of a symmetric $N \times N$ matrix $X$ whose
eigenvalues are $ x_1, x_2, \ldots, x_N $\,,
the external potential reads

$
V^{(N)}(x_1, x_2, \ldots, x_N) =
\mbox{det} X \mbox{tr}(X^{-2})\,.
$
}

\be\label{eq:superpot}
W^{(N)} = i \sqrt{x_1 x_2 \ldots x_N}\,.
\ee

Below it is useful to treat the internal degrees of
freedom entering into the Hamiltonian (\ref{eq:ECM})
in the Cartesian form
\be \label{eq:angmom}
l_{ab} = y_a \pi_b - y_b \pi_a\,,
\ee
where the internal variables $y_a$ and $\pi_a$ combine
the canonical pairs
with the canonical symplectic form.
The Hamiltonian  (\ref{eq:ECM}) has the following discrete
symmetries \cite{PolyDiscrete}:\\
\begin{itemize}
\item {Parity $P$}
\be
\left(
\begin{array}{c}
x_i  \\
p_i
\end{array}
\right)
\mapsto
\left(
\begin{array}{c}
-x_i  \\
-p_i
\end{array}
\right) \,,
\qquad
\left(
\begin{array}{c}
y_i  \\
\pi_i
\end{array}
\right)
\mapsto
\left(
\begin{array}{c}
-y_i  \\
-\pi_i
\end{array}
\right) \,,
\ee
\item{ Permutation symmetry $M$ }
\be
\left(
\begin{array}{c}
x_i  \\
p_i
\end{array}
\right)
\mapsto
\left(
\begin{array}{c}
x_{M(i)}  \\
p_{M(i)}
\end{array}
\right)\,,
\qquad
\left(
\begin{array}{c}
y_i  \\
\pi_i
\end{array}
\right)
\mapsto
\left(
\begin{array}{c}
y_{M(i)}  \\
\pi_{M(i)}
\end{array}
\right) \,,
\ee
\end{itemize}
where $M$ is the element of the permutation group $S_N$.
The  manifold of phase space defined as
\bea \label{eq:constraints1}
&& x_a + x_{N-a+1} = 0  \,, \quad p_a + p_{N-a+1} = 0  \,, \\ \label{eq:constraints2}
&& y_a + y_{N-a+1} = 0  \,, \quad \pi_a + \pi_{N-a+1} = 0
\eea
is invariant under the action of the symmetry group $z = D(z)$ where
\be
D = P \times M
\ee
and $M$ is specified as
$
M(a) = N - a + 1 \,.
$

In order to project onto the manifold  described  by
constraints (\ref{eq:constraints1})-(\ref{eq:constraints2})\,,
we use the Dirac method to deal with the second class constraints.
Let us introduce the Dirac brackets between the arbitrary functions $F$ and
$G$ of all variables $(x_a, p_a, y_a, \pi_a)$ as
\be
\{F \,, G\}_D =
\{F \,, G\} - \{F\,, Z_a\} \{Z_a\,, Z_b\}^{-1} \{Z_b\,, G\}
\ee
where $Z_a$ denote all second class constraints
$Z_a := (\chi_a, \Pi_a, \bar\chi_a, \bar\Pi_a)$, $a = 1, \cdots,\frac{N}{2}$
\bea  \label{CS1}
&& \chi_a = \frac{1}{\sqrt 2}
\left( x_a + x_{N-a+1} \right)\,, \qquad
{\bar\chi}_a = \frac{1}{\sqrt 2}
\left( y_a + y_{N-a+1} \right)\,,\\
&& \Pi_a = \frac{1}{\sqrt 2}  \label{CS2}
\left( p_a + p_{N-a+1} \right)\,, \qquad
{\bar\Pi}_a = \frac{1}{\sqrt 2}
\left( \pi_a + \pi_{N-a+1} \right)
\eea
with the canonical algebra
\bea
&& \{\chi_a , {\bar\chi}_b \} = \{ \Pi_a , {\bar\Pi}_b \} =
\{ \chi_a , {\bar\Pi}_b \} = \{ {\bar\chi}_a , \Pi_b \} = 0 \,, \\
&& \{ \chi_a , \Pi_b \} = \delta_{ab} \,,\quad \quad
\{ {\bar\chi}_a , {\bar\Pi}_b \} = \delta_{ab} \,.
\eea
Thus, the fundamental Dirac brackets are
\bea
\{x_a , p_b\}_D = \frac{1}{2} \delta_{ab}\,,\quad \quad
\{y_a , \pi_b\}_D = \frac{1}{2} \delta_{ab}\,.
\eea

After the introduction of these new brackets,
one can treat all constraints in the strong sense.
Letting the constraint functions vanish, the system with Hamiltonian (\ref{eq:ECM})
reduces to the following one
\be \label{eq:YMM}
H_{red} = \frac{1}{2} \sum_{a=1}^{\frac{N}{2}}  p^2_a +
\frac{1}{2}\sum_{a\not=b}^{\frac{N}{2}} l_{ab}^2 k_{ab}^2 +
\frac{g^2}{2} \  \sum_{a \not= b}^{\frac{N}{2}}  x_a^2 x_b^2 \,,
\ee
where
\be
k^2_{ab} = \frac{1}{(x_a +  x_b)^2} + \frac{1}{(x_a - x_b)^2}\,
\ee
Expression  (\ref{eq:YMM})
for  $N=6$
coincides with the Hamiltonian of the $SU(2)$ Yang-Mills mechanics
after taking  into account that after projection onto
the constraint shell (CS) (\ref{CS1})-(\ref{CS2})\,,
the potential (\ref{pot6}) reduces to the potential of Yang-Mills mechanics
\be
V^{(6)}(x_1,\cdots , x_6)_{\bigl\vert CS} =
\frac{1}{2} \left( x_1^2 x_2^2 + x_1^2 x_3^2 + x_2^2 x_3^2 \right)\,.
\ee


\section{Lax pair representation for Yang-Mills mechanics in zero coupling limit}


The conventional perturbative scheme of non-Abelian gauge theories starts
with the zero approximation of the free theory.
However, the limit of the zero coupling constant is not quite trivial.
If the coupling constant in the initial Yang-Mills action vanish,
the non-Abelian gauge symmetry reduces to the
$U(1)\times U(1)\times U(1)$ symmetry.
In this section,
we shall discuss this free theory limit for the case of the unconstrained Yang-Mills mechanics.
The solution of the  corresponding zero coupling limit of the
Yang-Mills mechanics in the form of a Lax representation will be given.
The relation between (\ref{eq:ECM}) and (\ref{eq:YMM}) allows one to construct
the Lax pair for the free part of the Hamiltonian (\ref{eq:YMM}) ($g=0$)
using the known Lax pair for the Euler-Calogero-Moser system (\ref{eq:ECM})
without an external potential term ($g=0$).

According to the work of S.Wojciechowski \cite{Wojciechowski}\,,
the Lax pair for the system with Hamiltonian
\be \label{eq:PECM}
H_{ECM} = \frac{1}{2} \sum_{a=1}^N  p^2_a +
\frac{1}{2}\sum_{a\not=b}^N \frac{l_{ab}^2}{(x_a - x_b)^2}
\ee
is
\bea
&& L_{ab} = p_a \delta_{ab} - (1 - \delta_{ab})
\frac{l_{ab}}{x_a - x_b}\,,\\
&& A_{ab}= (1 - \delta_{ab})\frac{l_{ab}}{(x_a - x_b)^2}\,.
\eea
and the equations of motion in Lax form are
\bea
\dot L & = & [A , L] \,,\\
\dot l & = & [A , l] \,,
\eea
where the matrix $ (l)_{ab} = l_{ab}$.

The introduction of Dirac brackets allows one to use the Lax pair
of higher dimensional Euler-Calogero-Moser model (namely $A_6$)
for the construction of Lax pairs ($L_{YMM}, A_{YMM}$)
of free Yang-Mills mechanics by performing the projection onto the
constraint shell (\ref{CS1})-(\ref{CS2})
\bea
L^{ECM}_{6\times 6} \vert_{CS} = L_{YMM}\,,\quad \quad
A^{ECM}_{6\times 6} \vert_{CS} = A_{YMM}\,.
\eea

Thus, the explicit form of the Lax pair matrices for the free $SU(2)$ Yang-Mills
mechanics is given by the following $ 6\times 6$ matrices


\bea \label{eq:Lmatrix}
\begin{array}{lr}
L_{YMM} =
\left (
\begin{array}{ccc|ccc}
           p_1            & -\frac{l_{12}}{x_1 - x_2} & -\frac{l_{13}}{x_1 - x_3}~~~ & \frac{l_{13}}{x_1 + x_3} & \frac{l_{12}}{x_1 + x_2} &    0  \\
-\frac{l_{12}}{x_1 - x_2} &            p_2            & -\frac{l_{23}}{x_2 - x_3}~~~ & \frac{l_{23}}{x_2 + x_3} &          0               & -\frac{l_{12}}{x_1 + x_2}  \\
-\frac{l_{13}}{x_1 - x_3} & -\frac{l_{23}}{x_2 - x_3} &             p_3~~~           &               0          & -\frac{l_{23}}{x_2 + x_3}& -\frac{l_{13}}{x_1 + x_3}  \\[0.2cm]
\hline
\frac{l_{13}}{x_1 + x_3} & \frac{l_{23}}{x_1 + x_2} &                0~~~            &          -  p_3          & -\frac{l_{23}}{x_2 - x_3} & -\frac{l_{13}}{x_1 - x_3} \\
\frac{l_{12}}{x_1 + x_2} &         0                & -\frac{l_{23}}{x_2 + x_3}~~~ & -\frac{l_{23}}{x_2 - x_3}   &          - p_2           & -\frac{l_{12}}{x_1 - x_2} \\
              0          & -\frac{l_{12}}{x_1 + x_2}& -\frac{l_{13}}{x_1 + x_3}~~~ & -\frac{l_{13}}{x_1 - x_3}   & -\frac{l_{12}}{x_1 - x_2}&          - p_1
\end{array}
\right )
\end{array}
\eea
and
\bea \label{eq:Amatrix}
\begin{array}{lr}
A_{YMM} =
\left(
\begin{array}{ccc|ccc}
             0                & \frac{l_{12}}{(x_1 - x_2)^2} & \frac{l_{13}}{(x_1 - x_3)^2}~~~ & -\frac{l_{13}}{(x_1 + x_3)^2} & -\frac{l_{12}}{(x_1 + x_2)^2} &          0                 \\
-\frac{l_{12}}{(x_1 - x_2)^2} &              0               & \frac{l_{23}}{(x_2 - x_3)^2}~~~ &-\frac{l_{23}}{(x_2 + x_3)^2}  &              0                & \frac{l_{12}}{(x_1 + x_2)^2}\\
-\frac{l_{13}}{(x_1 - x_3)^2} & -\frac{l_{23}}{(x_2 - x_3)^2}&                 0~~~            &               0               & \frac{l_{23}}{(x_2 + x_3)^2}  & \frac{l_{13}}{(x_1 + x_3)^2} \\[0.2cm]
\hline
\frac{l_{13}}{(x_1 + x_3)^2}  & \frac{l_{23}}{(x_1 + x_2)^2} &                 0~~~            &               0               & -\frac{l_{23}}{(x_2 - x_3)^2} & -\frac{l_{13}}{(x_1 - x_3)^2} \\
\frac{l_{12}}{(x_1 + x_2)^2}  &              0               & -\frac{l_{23}}{(x_2 + x_3)^2}~~~& \frac{l_{23}}{(x_2 - x_3)^2}  &              0                & -\frac{l_{12}}{(x_1 - x_2)^2} \\
             0                & -\frac{l_{12}}{(x_1 + x_2)^2}& -\frac{l_{13}}{(x_1 + x_3)^2}~~~& \frac{l_{13}}{(x_1 - x_3)^2}  & \frac{l_{12}}{(x_1 - x_2)^2}  &           0
\end{array}
\right )
\end{array}
\eea

The equations of motion for the $SU(2)$ Yang-Mills mechanics in the zero constant coupling limit
read in a Lax form as
\bea
\dot L_{YMM} & = & [A_{YMM} , L_{YMM}] \,,\\
\dot l_{YMM} & = & [A_{YMM} , l_{YMM}] \,,
\eea
where the matrix $l_{YMM}$ is

\bea \label{eq:lmatrix}
\begin{array}{lr}
l_{YMM} =
\left(
\begin{array}{ccc|ccc}
    0   & l_{12} & l_{13}~~~ &-l_{13} &-l_{12} &  0   \\
-l_{12} &   0    & l_{23}~~~ &-l_{23} &    0   & l_{12}\\
-l_{13} &-l_{23} &      0~~~ &   0    & l_{23} & l_{13} \\[0.2cm]
\hline
 l_{13} & l_{23} &      0~~~ &   0    &-l_{23} & -l_{13} \\
 l_{12} &    0   &-l_{23}~~~ & l_{23} &   0    & -l_{12} \\
   0    & -l_{12}&-l_{13}~~~ & l_{13} & l_{12} &    0  \
\end{array}
\right )\,.
\end{array}
\eea


\section{Acknowledgments}


We are grateful to V.I. Inozemtsev, M.D. Mateev and H-P. Pavel for discussions.
We would like to thank E. Langmann for useful comments on the relations
between 2-dimensional Yang-Mills theory on a cylinder and
Calogero-Moser systems.
B.G. Dimitrov is also acknowledged for reading of the manuscript.
The work of A.M.K. was supported in part by the Russian
Foundation for Basic Research under grant No. 96-01-00101.


\end{document}